# Magnetostriction of $AlFe_2B_2$ in High Magnetic Fields

S. Sharma,[1] A. E. Kovalev,[1] D. J. Rebar[1], D. Mann,[2] V. Yannello,[3] M. Shatruk,[2] A.V. Suslov,[1] J. H. Smith,[1] T. Siegrist[1,4]

[1] *National High Magnetic Field Laboratory, Tallahassee, FL 32310, USA*
[2] *Department of Chemistry and Biochemistry, Florida State University, Tallahassee, FL 32306, USA*
[3] *Department of Chemistry, Biochemistry and Physics, University of Tampa, Tampa, FL 33606, USA*
[4] *Department of Chemical and Biomedical Engineering, FAMU-FSU College of Engineering, Tallahassee, FL 32310, USA*

**Abstract:**

Using the experimental capability of the novel X-ray diffraction instrument available at the 25 Tesla Florida Split Coil Magnet at the NHMFL, Tallahassee we present an extensive investigation on the magnetostriction of polycrystalline $AlFe_2B_2$. The magnetostriction was measured near the ferromagnetic transition temperature (Curie temperature $T_C$ = 280 K, determined via *DC* magnetization measurements), namely, at 250, 290, and 300 K. $AlFe_2B_2$ exhibits an anisotropic change in lattice parameters as a function of magnetic field near the Curie temperature, and a monotonic variation as a function of applied field has been observed, i.e., the *c*-axis increases significantly while the *a*- and *b*-axes decrease with the increasing field in the vicinity of $T_C$, irrespective of the measurement temperature. The volume magnetostriction decreases with decreasing temperature and changes its sign across $T_C$. Density functional theory calculations for the non-polarized and spin-polarized (ferromagnetic) models confirm that the observed changes in lattice parameters due to spin polarization are consistent with the experiment. The relationships for magnetostriction are estimated based on a simplified Landau model that agrees well with the experimental results.

## Introduction

Magnetostructural or magnetoelastic coupling is a strong coupling between magnetic and structural responses observed, for example, in magnetoelectric multiferroics (type II)[1,2]. It is the common driving mechanism responsible for the use of a material in magnetomechanical devices[3,4], magnetic cooling/refrigeration[5,6]. The materials exhibiting magnetostructural coupling demonstrate a range of interesting behaviors, including magnetic shape memory effect[7], magnetocaloric effect[8,9], magnetostriction or magnetic field induced strain[10–12], and very large magnetoresistance[13]. Recently, $AlFe_2B_2$ has gathered considerable attention due to its promising magnetocaloric properties near room temperature[14–17]. Although the change in entropy with magnetic field (H) in this intermetallic compound is moderate when compared to state-of-the-art magnetocaloric materials, such as $Gd_5Si_4$ and related systems[8,9,18–21], the inexpensive earth-abundant elements and straightforward synthesis make $AlFe_2B_2$ a promising candidate for magnetocaloric applications. The typical value of the isothermal entropy change is 4.1 J/(kg·K) at 2 T and 7.7 J/(kg·K) at 5 T[14]. The crystal structure of $AlFe_2B_2$ was first reported by Jeitschko[22], and the ferromagnetic (FM) transition temperature ($T_C$) was reported to vary between 274 K and 320 K depending on the synthesis conditions[14–16,23–28]. It has been recently shown that the variation in $T_C$ stems from a narrow stoichiometry range, $Al_{1-y}Fe_{1+y}B_2$ ($-0.01 \leq y \leq 0.01$), with higher $T_C$ values observed at the smaller Al/Fe ratios[15]. Neutron diffraction studies showed that the magnetic moments are aligned along the *a*-axis in the FM state[29], while density functional theory (DFT) predicted the moments to be in the *ab*-plane[30].

Recently, Ke *et al.*[30] have studied the electronic structure and magnetic response of $AlT_2B_2$ (T = Fe, Mn, Cr, Co, Ni) using DFT and suggested that the magnetization is strongly affected by the change in the lattice parameter *c*, which is perpendicular to the zigzag chains of B atoms and lies in plane with the [$T_2B_2$] layers that are parallel to the *ac* plane (Fig. 1). Consistent with theoretical predictions, Lejeune *et al.*[31] have confirmed that it is indeed the change in the *c*-axis length and associated $(Fe\text{-}Fe)_{c\text{-axis}}$ interatomic distance that has the largest effect on $T_C$, while $T_C$ is mostly independent of the ($b/a$) ratio, indicating the negligible role of the *a*- or *b*- axis in affecting $T_C$. The recent detailed study of magnetic properties of single-crystalline $AlFe_2B_2$ suggested itinerant magnetic behavior, based on the Rhodes-Wohlfarth ratio of ~1.14 [23]. Several groups have investigated the effect of alloying Mn, Cr, Co, or Ni on the Fe site and C substitution on the B site[30,31]. The effect of pressure has also been studied, demonstrating that $T_C$ is suppressed by ~19 K at a pressure of 2.24 GPa[23]. The magnetocrystalline anisotropy field was reported to be 1 T along the *b*-axis and 5 T along the *c*-axis, consistent with the DFT results[30]. Temperature dependent X-ray diffraction (XRD) results on $AlFe_2B_2$ show that both the *a* and *b*- axes increase while *c* decreases when cooling the sample from 298 K to 200 K[32].

Despite several reports suggesting a strong correlation between magnetic and structural properties[25,30–32], the direct changes in the crystal structure of $AlFe_2B_2$ under external magnetic fields have not been investigated. A possible reason for this gap is the lack of non-trivial experimental setups where both temperature and magnetic field can be varied in a broad range to investigate the evolution of structural properties across the magnetic phase transition. Here, we report our experimental study of magnetostriction behavior in $AlFe_2B_2$ near the FM ordering temperature in magnetic fields up to 25 T. Our results provide detailed insights into the structural changes of $AlFe_2B_2$ across $T_C$ and highlight the experimental capabilities of the novel high magnetic field XRD setup used for the present work. The observed magnetoelastic coupling is analyzed using Landau theory and spin polarized DFT calculations .

**Experimental Details**

The sample has been synthesized using arc melting, with the detailed procedure described previously[14]. Briefly, a mixture of starting materials in the Al:Fe:B = 3:2:2 ratio, with a total mass of 0.35 g, was pressed into a pellet, arc-melted, and subjected to annealing at 900 °C for 1 week. The $Al_{13}Fe_4$ byproduct was removed by washing the sample with dilute hydrochloric acid (1:1 v/v). The sample purity was checked by powder XRD which confirms the single phase nature of the sample[14].

To investigate the magneto-elastic effect in $AlFe_2B_2$, we used a custom diffraction setup integrated with the Florida Split Coil Magnet at the National High Magnetic Field Laboratory (NHMFL) and capable of diffraction in the presence of high *DC* magnetic field of up to ±25 T [33]. To access the sample space, the magnet has four identical optical ports defining an angular diffraction range of 45° in the forward direction. Higher diffraction angles are available through side ports as described previously[33]. The Mo Kα radiation is generated by a rotating anode source with a maximum power of 18 kW, either Zr-filtered (10 μm) or reflected off a multilayer mirror to provide a monochromatic Mo Kα radiation spectrum. A Pilatus 300 K-W X™ hybrid pixel detector, customized to tolerate the magnetic fringe field of the split coil magnet, is used to detect the X-rays at a distance of approximately 1200 mm from the sample. The detector was mounted on a VELMEX™ linear slide on an optical table near the X-ray beam exit window to access a wider range of diffraction angles [33]. The DAWN software[34], developed at the DIAMOND synchrotron, has been employed to convert the detector images to 2θ – intensity data based on geometrical calibration parameters obtained using a NIST SRM 660b $LaB_6$ reference sample[35]. JANA2006[36] has been used to LeBail fit the data in order to get the field and temperature dependences of the unit cell parameters. Since the measurements presented here on $AlFe_2B_2$ involve a high *DC* magnetic field (H) of 25 T, the instrument was also calibrated with $LaB_6$ under the same diffraction condition, temperature and magnetic field, in order to avoid influencing the data analysis by any effect of magnetic fringe fields[33,37]. The results on the $LaB_6$ sample are given in the supplementary information (Fig. S1), together with additional details of the diffraction system[37]. *DC* magnetization measurements have been performed as a function of temperature and magnetic field to produce an Arrott plot to determine the $T_C$ for the studied sample using a SQUID magnetometer[37].

DFT calculations were accomplished using the Vienna ab initio Simulation Package (VASP)[38]. Published structural parameters of $AlFe_2B_2$[22] were used for the initial structural geometry, which was subsequently optimized with and without inclusion of spin polarization. PAW-PBE pseudopotentials were used for all elements.

## Results and Discussion

Our sample of $AlFe_2B_2$ crystallizes in the orthorhombic space group *Cmmm* and undergoes FM ordering at $T_C$ = 280 K with a saturation magnetization value of nearly 2.50 $\mu_B$/f.u. Further details of the *DC* magnetization measurements are given in the supplementary information (Fig. S2 and S3)[37]. Figure 1 shows a representation of the unit cell containing two $AlFe_2B_2$ formula units: Layers of Al atoms alternate with the $Fe_2B_2$ layers along the *b*-axis. The boron atoms form zig-zag chains that run along the *a*-axis while the Fe atoms connect these chains in the *ac*-plane (Fig. 1c). The *bc*-plane also reveals linear chains of Fe atoms along the *c*-axis. The nearest Fe-Fe distance is equal to the *c* parameter.

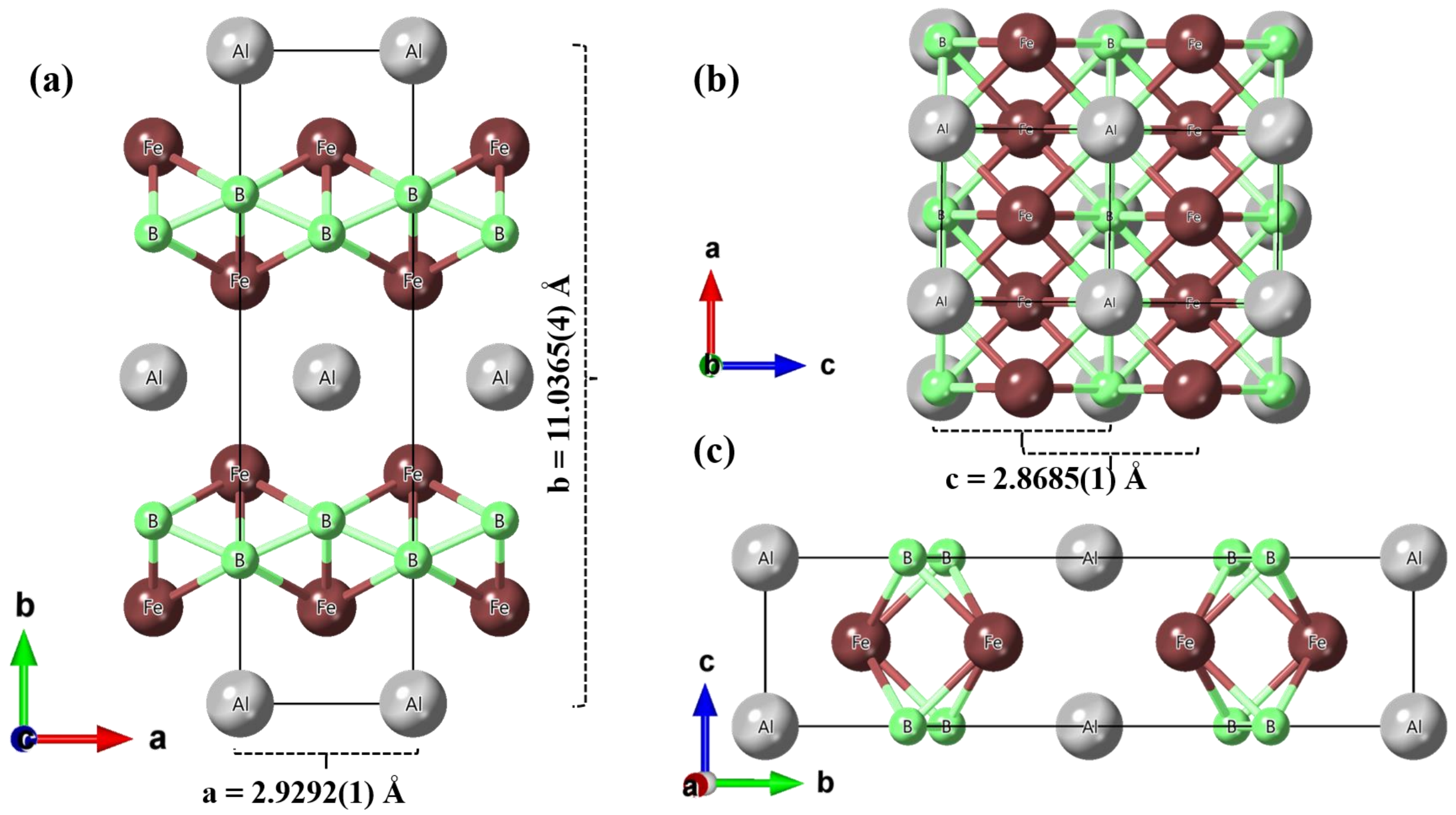

**Fig. 1:** Unit cell representation of $AlFe_2B_2$ crystal structure in (a) *ab*-, (b) *ac*-, and (c) *bc*-planes. The unit cell parameters indicated in the drawing were obtained by fitting an XRD pattern collected at 300 K and zero applied magnetic field. (Silver spheres: Aluminum, Green spheres: Boron and Brown spheres: Iron atoms).

To determine the magnetostrictive or magnetoelastic interactions derived from the FM exchange coupling between Fe moments, we carried out XRD measurements as a function of applied magnetic field at temperature of 300, 290, and 250 K. A thin layer of a powdered sample was placed on a copper substrate oriented parallel to the magnetic field. The XRD patterns at 300 K were recorded in magnetic fields of 0, 25, and –25 T. At 290 and 250 K, data were collected at 0, 4, 7, 10, 15, 20, and 25 T, and at 250 K, reversed magnetic fields (up to -25 T, not shown here) were also included. Fig. 2(a-c) show the XRD line profiles of the (130), (060), and (041) reflections measured at 300 K and 0, 25, and –25 T. Clear shifts in the peak positions of these reflections are observed. The (130) and (060) reflections shift towards higher 2θ values while the (041) reflections shifts toward lower 2θ values with increasing magnetic field, indicating that the lattice parameters *a* and *b* both decrease while *c* increases with the field.

To extract a more precise field dependence of the lattice parameters, LeBail fitting of several peaks was carried out. The magnitude of the shift is more pronounced for the (041) reflection as compared to the other two reflections, indicating that the applied magnetic field affects the *c* parameter more than the *a* and *b* parameters, consistent with DFT results which are discussed in the later section. The refined values of the lattice parameters at 300 K are *a* = 2.9292(1) Å, *b* = 11.0365(4) Å, *c* = 2.8685(1) Å in the absence of a magnetic field and *a* = 2.9277(1) Å, *b* = 11.0300(5) Å, *c* = 2.8736(1) Å at 25 T.

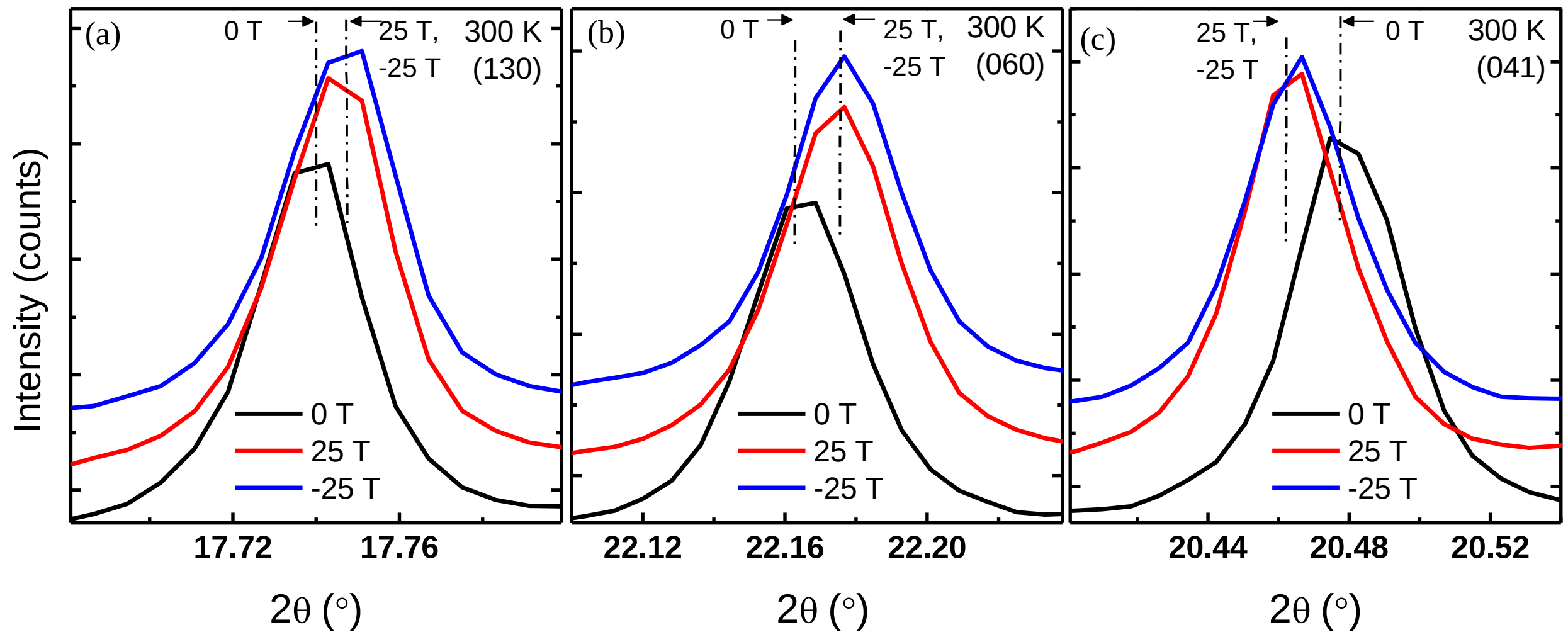

Fig. 2: XRD peak profiles of (a) (130), (b) (060), and (c) (041) reflections recorded at 300 K under 0, 25, and –25 T applied field.

To investigate the effect of the magnetic field on the $AlFe_2B_2$ lattice across $T_C$ = 280 K , XRD patterns were collected above (290 K) and below (250 K) $T_C$. Figs. 3(a-c) show the field-dependent XRD reflection profiles of the (130), (060), and (041) reflections at 290 K, while Figs. 3(d-f) show the same peak profiles at 250 K. Again, significant angular shifts are observed for the (041) reflection at 290 K and 250 K, with a smaller magnitude at 250 K than at 290 K.

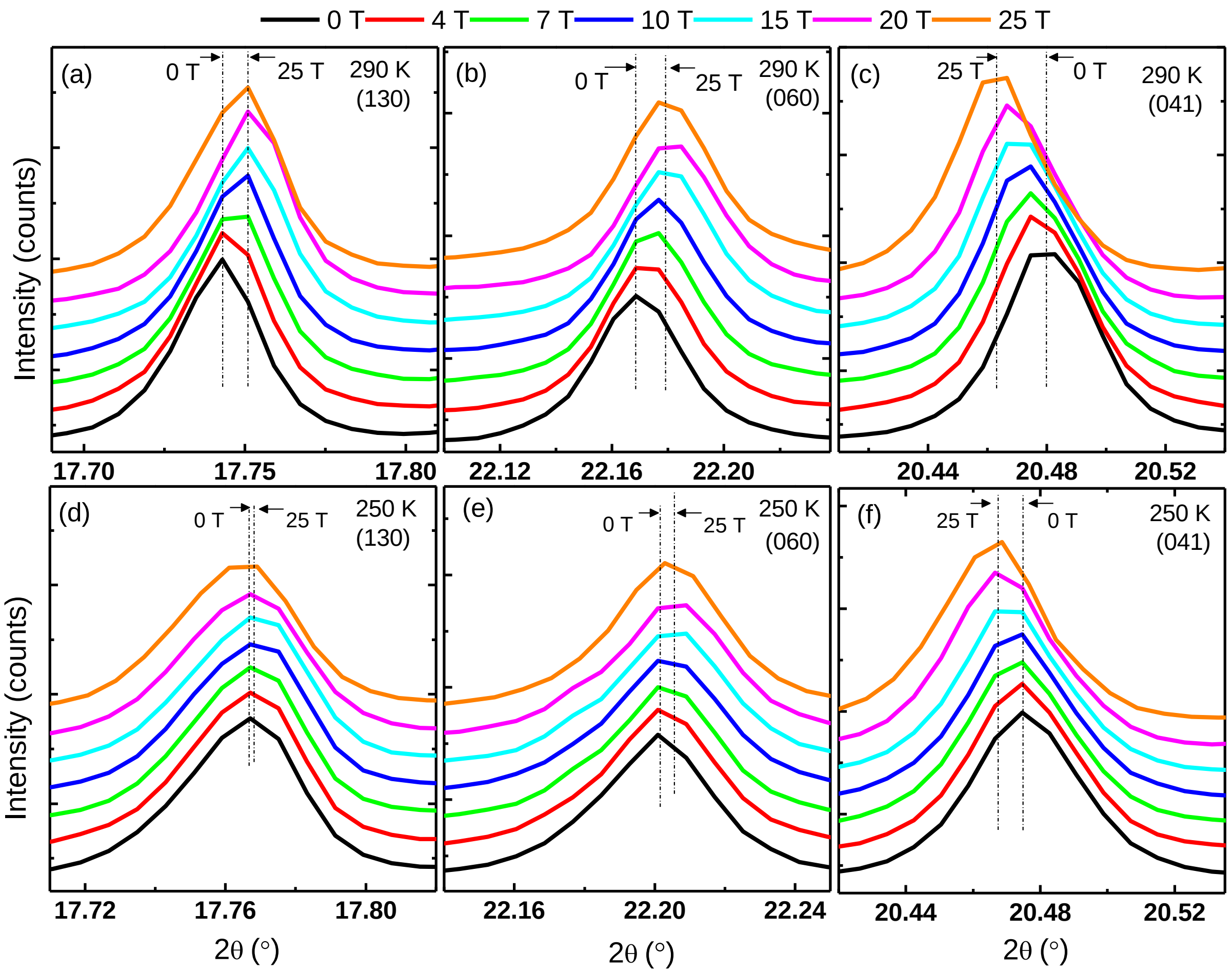

**Fig. 3:** XRD reflection profiles of (130), (060), and (041) reflections recorded at (a-c) 290 K and (d-f) 250 K as a function of applied magnetic field. The dashed vertical lines in each panel show the shift in the peak position at 25 T with respect to the signal recorded in the absence of field.

To confirm these field-dependent variations in lattice parameters at 290 K and 250 K, the full XRD patterns were subjected to LeBail fitting. For both temperatures, the *c*-axis increases while the *a*- and *b*-axes decrease with magnetic field, as seen in Fig. 4(a,b). Consequently, the anisotropic strain is positive along *c* and negative along *a* and *b*, as shown in Figs. 5 (a-c), with the magnitude of strain maximal along the *c*-axis and minimal along the *a*-axis.

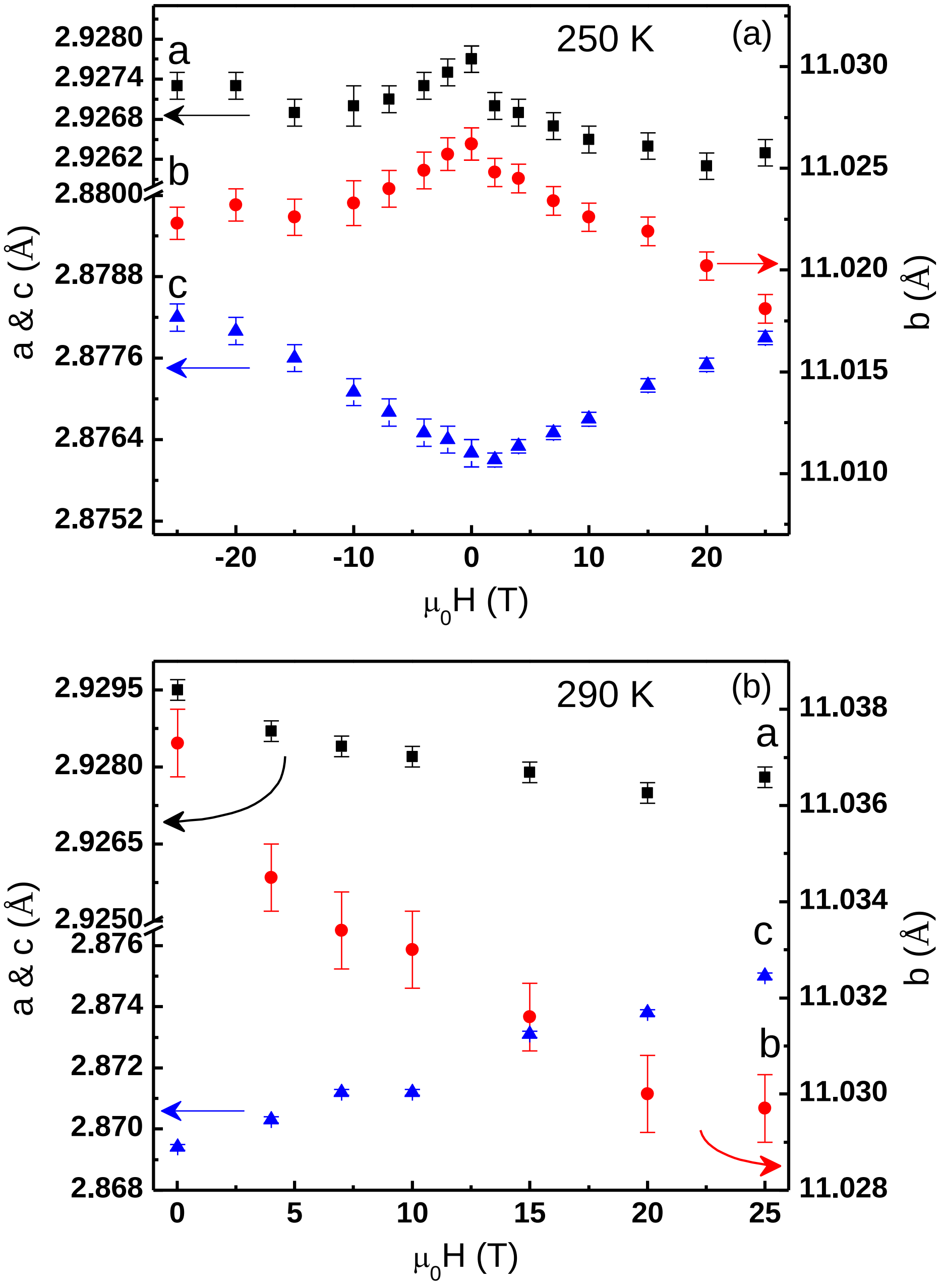

**Fig. 4:** The field dependence of lattice parameters at (a) 250 K and (b) 290 K. Black squares: *a*-axis, red circles: *b*-axis, blue triangles: *c*-axis.

Interestingly, the change in the *c*-axis length is more pronounced with temperature than the corresponding changes in the *a*- or *b*-axis. The overall magnitude of change in the lattice parameters with magnetic field is larger at 290 K than at 300 K and 250 K, but the overall volume remains nearly constant at 300 K and 290 K while it slightly decreases at 250 K. Therefore, near $T_C$, the increase in the *c*-axis is compensated by decreases in the *a*- and *b*- axes.

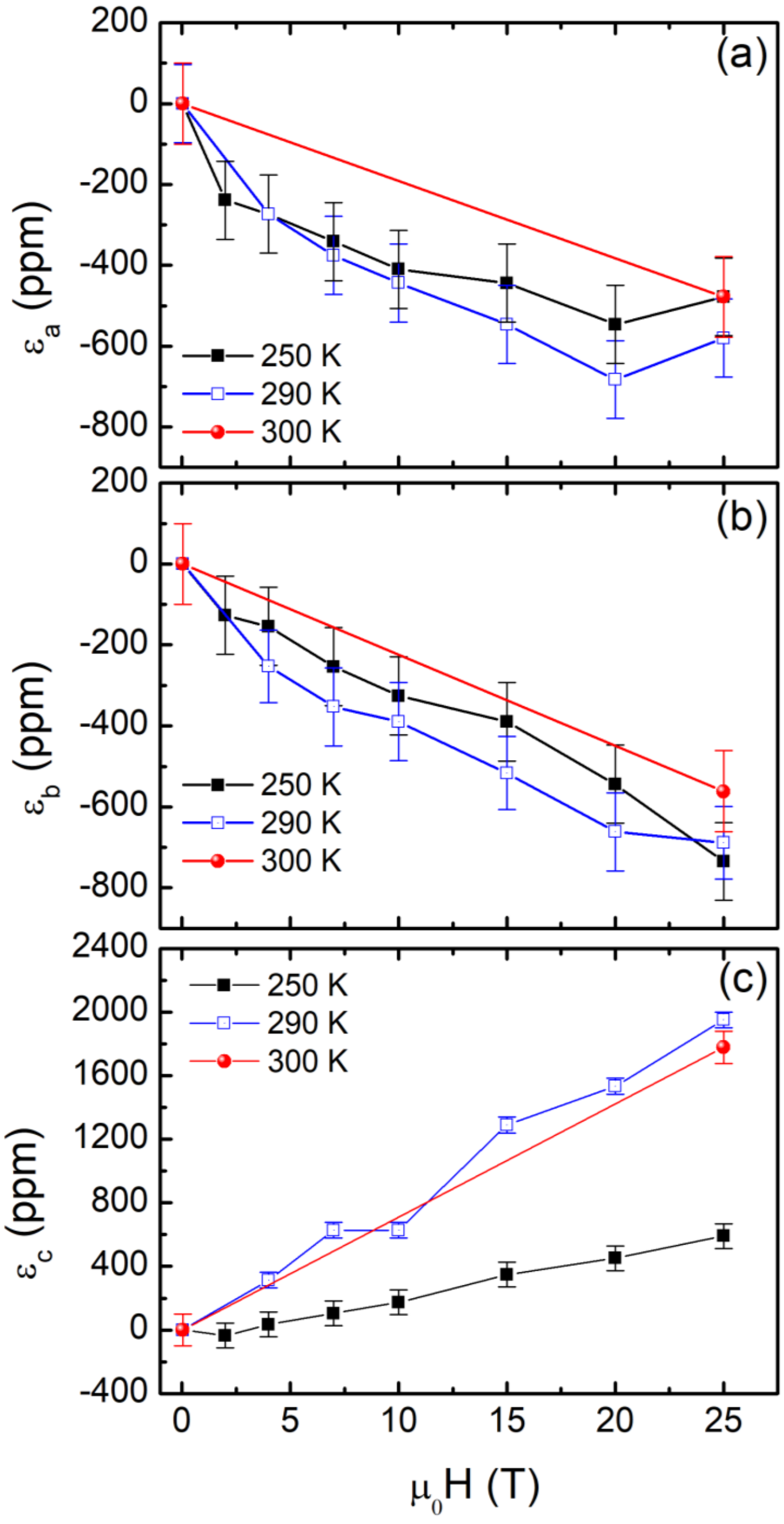

**Fig. 5:** The anisotropic magnetic field induced strain measured along (a) *a*- (b) *b*- and (c) *c*-axis at 250, 290 and 300 K. The strain is positive in the *c*-direction while it is negative along the *a*- and *b*-direction.

Furthermore, it is interesting to compare the order of magnitude of the effect of temperature and magnetic field on lattice parameters per degree and per tesla, respectively. These effects are similar, and the details are discussed in the supplementary information[37].

We cannot measure the exact volume magnetostriction in our experimental setup because the lattice strains are measured at different field orientations with respect to the crystallographic axes. We believe, however, that it is still informative to evaluate the volume magnetostriction. The evolution of such effective volume magnetostriction with field is dependent on temperature, as shown in Fig. 6. The typical value of the effective volume magnetostriction (ΔV/V, %) at 300, 290 K, and 250 K and 25 T are 0.070, 0.068, and -0.062%, respectively. The negative value observed at 250 K is mostly due to the reduced effect of the applied field on the *c*-axis.

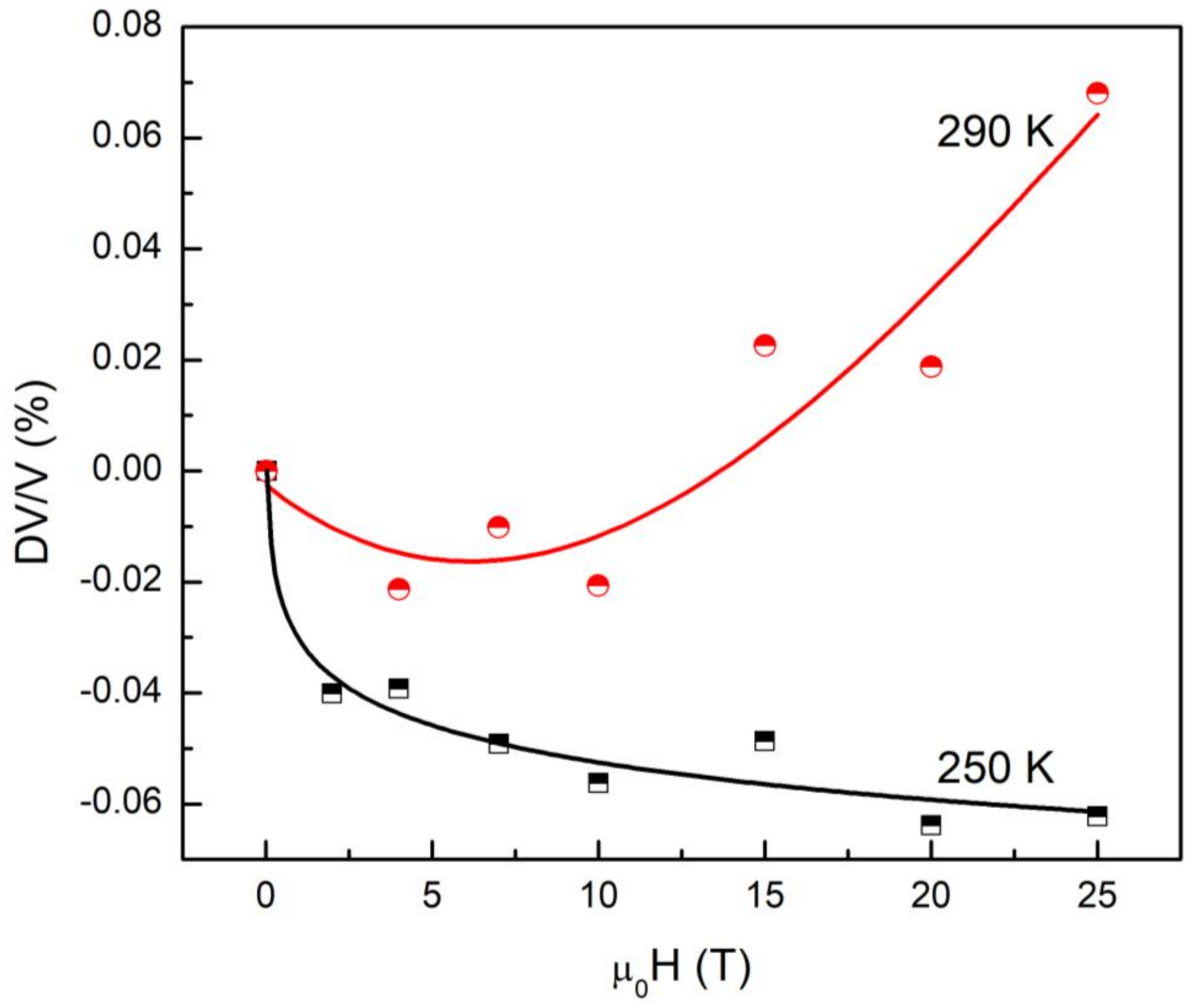

**Fig. 6:** Volume magnetostriction measured at 290 K and 250 K as a function of applied magnetic field. The lines are the guide to the eyes.

The reason for observing a larger change in *c* parameter at 290 and 300 K as compared to the change at 250 K is as follows: Above $T_C$, the thermal fluctuations will oppose the effect of the magnetic field on the lattice by counteracting the spin alignment, and below $T_C$, with ordered spins, the effect of the magnetic field on the lattice will be reduced. Thus, the largest magnetostriction effect can be expected near $T_C$, and it should decrease on both sides of the transition, due to thermal fluctuations above $T_C$ and due to spin order below $T_C$.

To compare the magnetic energy and the strain energy at 25 T, we estimate both in the following way: with the saturation magnetization of 2.06 $\mu_B$/f.u. observed for our sample at 5 T, the magnetic energy (-MH) is of the order of ~57.5 J/mol at 5 T and ~287.6 J/mol at 25 T, assuming saturation at 5 T. Using the bulk modulus of 213.42 GPa[39], the density of 5.75 g/cm$^3$ [23], and the experimentally observed value of strain ($\Delta c/c$) along the $c$-direction at 300 K, the elastic energy is estimated to be about 9.42 J/mol. This indicates that, even at 5 T, the magnetic energy is significantly larger than the elastic energy in $AlFe_2B_2$.

To further evaluate the changes in structural properties under applied magnetic field, we can consider the effect of the magnetic field within the framework of the Landau model for the phase transitions[40]. The simplified free energy per unit volume near the transition temperature can be written as[41]:

$$f = a(T - T_C)M^2 + \frac{bM^4}{2} - MB + \lambda\varepsilon M^2 + \frac{1}{2}C_{EL}\varepsilon^2 \qquad (1)$$

where $B$ is the external magnetic field, $M$ the sample magnetization , $\varepsilon$ is the strain, and $C_{EL}$ is the elastic constant. The fourth term ($\lambda\varepsilon M^2$) is the magnetoelastic energy (coupling the strain and magnetization) at the lowest order and the fifth term ($\frac{1}{2}C_{EL}\varepsilon^2$) is the elastic energy contribution to the Gibbs free energy. The magnetostriction is obtained by minimizing Equation (1) with respect to the strain:

$$\varepsilon = -\frac{\lambda M^2}{C_{EL}} = NM^2 \qquad (2)$$

With N = - ($\lambda$ /$C_{El}$), a magnetostriction constant. In $AlFe_2B_2$, the magnetostriction is anisotropic, so $N$ is a function of both the angle and the orientation of the magnetization. If the magnetoelastic energy is small in comparison to the first three terms in Equation (1), then $M$ can be calculated by ignoring the last two terms, and the anisotropic magnetostriction terms are obtained from Equation 2. The field dependence of the magnetization can be calculated from the Weiss mean field model[42] :

$$M = ng\mu_B S B_S\left(Sg\frac{\mu_B B + \mu_0\gamma\mu_B M}{k_B T}\right) = M_S B_S\left(\frac{\mu_B S g B}{k_B T} + \frac{3S}{S+1}\,\frac{T_C}{T}\frac{M}{M_S}\right) \qquad (3)$$

wh

We can compare the magnetoelastic coefficient with the jump in the thermal expansion coefficient at $T_C$. From our temperature-dependent data at zero field, we estimate this jump in the thermal expansion coefficient for the $b$-axis to be about $(1.6 \pm 0.1) \times 10^{-5}$K$^{-1}$. This value is of the same order as $4.4 \times 10^{-5}$K$^{-1}$ that can be estimated from the data by Oey *et al.*[32]. On the other hand, the jump in the thermal expansion coefficient is due to the appearance of the spontaneous magnetic moment below $T_C$, which will lead to magnetostriction according to Equation **Error! Reference source not found.**). The magnetic moment below the transition temperature can be estimated by using the Taylor expansion of the Brillouin function for S=1/2 near $T_c$, at zero magnetic field: $M^2 \approx 3M_s^2(1 - T/T_c)$. If this magnetostriction constant is used, the jump in the thermal expansion is about $4 \times 10^{-5}$ K$^{-1}$, which is close to the observed value. The sign of the effect is in agreement with the experimental results: the *b*-axis contracts if there is a spontaneous magnetization perpendicular to it.

To understand the influence of ferromagnetic ordering on the lattice parameters of $AlFe_2B_2$, we also performed DFT calculations on the non-polarized and spin-polarized (ferromagnetic) models, starting with the experimentally determined structure[22]. The unit cell parameters obtained after geometry optimization are listed in Table 1. As can be seen from these results, spin polarization has a minor effect on the $a$-axis, which contracts only slightly, while a somewhat larger contraction is observed for the $b$-axis. The largest change is seen in the $c$-axis, which is elongated by more than 5% upon spin polarization. The results of our calculations are in good agreement with the changes in the unit cell parameters calculated by Ke *et al.*[30], and they agree with the experimental observation of the relative changes in the unit cell parameters upon application of high magnetic field near the $T_C$ (Fig. 4).

Table 1. Results of geometry optimization for $AlFe_2B_2$ in the non-polarized and spin-polarized (ferromagnetic) models.

| **Parameter** | **Non-Polarized** | **Spin-Polarized** | **Relative Change** |
|---|---|---|---|
| ***a*** **(Å)** | 2.9297 | 2.9153 | 0.49% |
| ***b*** **(Å)** | 11.3485 | 11.0247 | –2.94% |
| ***c*** **(Å)** | 2.69676 | 2.8487 | 5.33% |

## Conclusion

$AlFe_2B_2$ exhibits anisotropic magnetostriction in an applied *DC* magnetic field up to 25 T. The unit cell parameter *c* increases while *a* and *b* decrease with increasing magnetic field, with the largest effect for the elongation of the *c*-axis, consistent with DFT calculations. Close to $T_C$, at 300 K and 290 K, the magnitude of the magnetostriction is larger than at 250 K. Furthermore, the fourth order magnetoelastic energy terms in magnetization should be comparable to the quadratic terms. The model based on Landau theory including quartic terms gives qualitative good agreement with the observed behavior of $AlFe_2B_2$ in high magnetic fields. The model correctly predicts that the magnetostrictive effects are largest in the vicinity of $T_C$ and drop off for higher and lower temperatures. While not all tensor components of the magnetoelastic tensor can be determined from powder diffraction measurements in high magnetic fields, the novel X-ray diffractometer installed in Florida Split Coil 25 T Magnet at the NHMFL has been instrumental in assessing the model for magnetostriction based on Landau theory. The main elements of the magnetoelastic coupling tensor could be determined, giving values for the *a*-axis as $-2.1\times10^{-6}$ $MPa^{-1}$, the *b*-axis as $-2.4\times10^{-6}$ $MPa^{-1}$, and $7.1\times10^{-6}$ $MPa^{-1}$for the *c*-axis (Details are provided in the supplementary information[37]). While the magnetostriction effects on the a- and b-axes are not strongly dependent on temperature, the magnetostriction is reduced at lower temperature along the c-axis, resulting in an overall negative volume magnetostriction at 250 K. The results of DFT calculations support the observed anisotropic changes in the lattice parameters of $AlFe_2B_2$ caused by ferromagnetic alignment of Fe moments.

## Acknowledgments

This work is supported by the National Science Foundation under award DMR – 1625780. Part of the work was carried out at the National High Magnetic Field Laboratory, which is supported by the National Science Foundation under Cooperative Agreement No. DMR-1644779 and the State of Florida.

## Supplementary Information

### Calibration of the diffractometer

We measured the $LaB_6$ diffraction pattern using NIST Standard Reference Material 660b[35] at room temperature and 17 K. The diffraction patterns have been measured utilizing a Mo $K_\alpha$ source. A multilayer graded mirror assembly have been used to align the X-rays. A Pilatus 300 K-W X™ hybrid pixel detector, customized to tolerate the magnetic fringe field of the split coil magnet, is used to detect the X-rays at a distance of approximately 1200 mm from the sample. The estimated FWHM for the (200) reflection at 17.07° is of the order of 0.015° which is roughly of the order of 1-2 pixel, allowing separation of the $K\alpha_1$ / $K\alpha_2$ radiation. The FWHM follows a Cagliotti function, with values U = 0 , V= 0 , and W = 0.33, indicating the diffractometer resolution in the forward scattering region is of the order of 0.015°. We have also tried to refine the U and V parameter but it results to the value with very large errors. A Pseudo-Voight peak shape function has been used to fit the peak profiles. The detector was mounted on a VELMEX™ slide on an optical table near the X-ray beam exit window at 1200 mm distance, enabled for linear translation to access a wider range of diffraction angles. With a pixel size of 172 μm and a sample to detector distance of 900 mm, a single pixel will span approximately 0.011 degree, indicating that the setup is close to the possible resolution limit. It is therefore possible to observed changes in the reflection position of the order of 0.01 degrees, resulting in an observable relative change in d-spacings of the order of $10^{-3}$.

Stability in magnetic fields of +25 T and -25 T Tesla at room temperature and 17 K was tested, with no shifts in the observed peaks due to the magnetic field. Here we are presenting the LaB6 data recorded at 17 K from -25 T and 25 T to determine any magnetic field effects on different parts of the diffractometer. Fig. S1(a) shows the profile fit XRD pattern of $LaB_6$ obtained at room temperature in the absence of a magnetic field. All the main peaks could be LeBail fitted with the cubic space group *Pm*-3*m*. The enlarged view in the inset shows the $K\alpha_1$ and $K\alpha_2$ of the (210) reflection and the corresponding LeBail fit. The refined lattice parameter of **a** = 4.1570(1)

Å agrees well with the value reported for $LaB_6$ 660b[34]. The rectangular boxed regions are excluded during the fitting as they contain the regions between the active detector modules. The sample was subsequently cooled down to 17 K and the XRD pattern was recorded with field values in the following sequence: 0, 10 T, 25 T, 0, -10 T and -25 T. Fig. S1(b) represents a stack of the XRD peak profiles of the (110) reflections ($K\alpha_1$ and $K\alpha_2$) measured at 17 K as a function of the applied magnetic field. The peak positions remain unchanged in field H, which confirms that the magnetic field has no appreciable effect on the XRD instrument used in the present work and therefore the magnetic field dependent effects observed are due to the specimen.

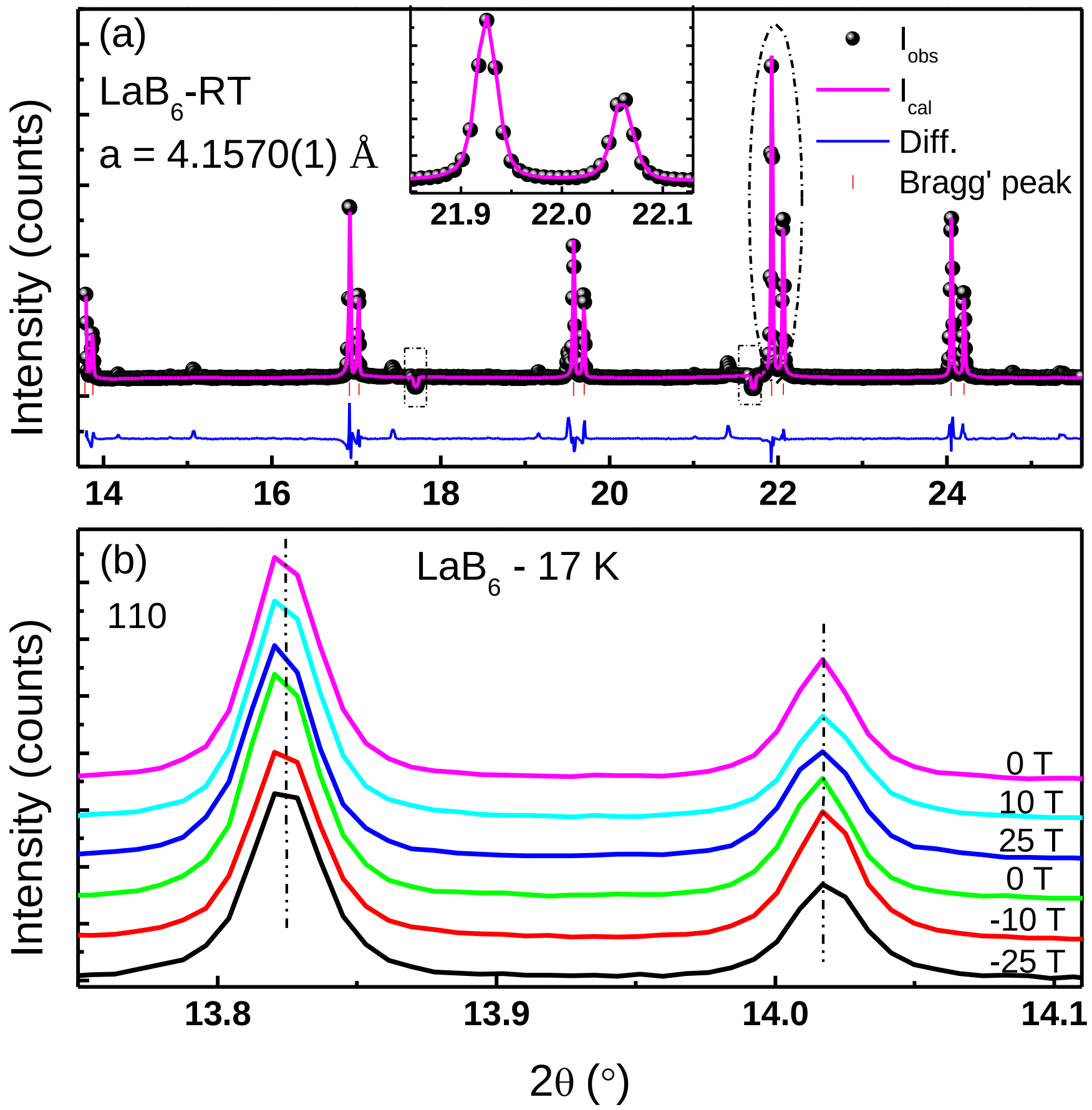

Fig. S1: (a) Le Bail fit XRD pattern of $LaB_6$ recorded at room temperature using the space group *Pm*-3*m*. The dashed rectangular boxed regions are the angular ranges which were excluded during the fitting. The inset shows an enlarged view of (210) reflection (highlighted in the encircled region). (b) Stack of (110) peaks ($K\alpha_1$ and $K\alpha_2$) recorded at 17 K under various applied field values are shown. No appreciable shift in the peak position was found in the applied magnetic field.

## DC magnetization behavior

Fig. S2(a,b) represents the *DC* magnetization behavior of $AlFe_2B_2$ sample measured as a function of temperature and magnetic field to obtain the onset of ferromagnetic ordering and to determine the saturation magnetization of the studied sample. The magnetization increases sharply when the sample is cooled below 300 K, exhibiting a dip near 280 K in the first order derivative which is shown in the inset of Fig. S2(b). To establish the value of $T_C$, Arrott plots are presented in Fig. S3 which give $T_C$ approximately 280 K, consistent with the literature[14–16,25]. The saturation moment is near 2.5 $\mu_B$/f.u.

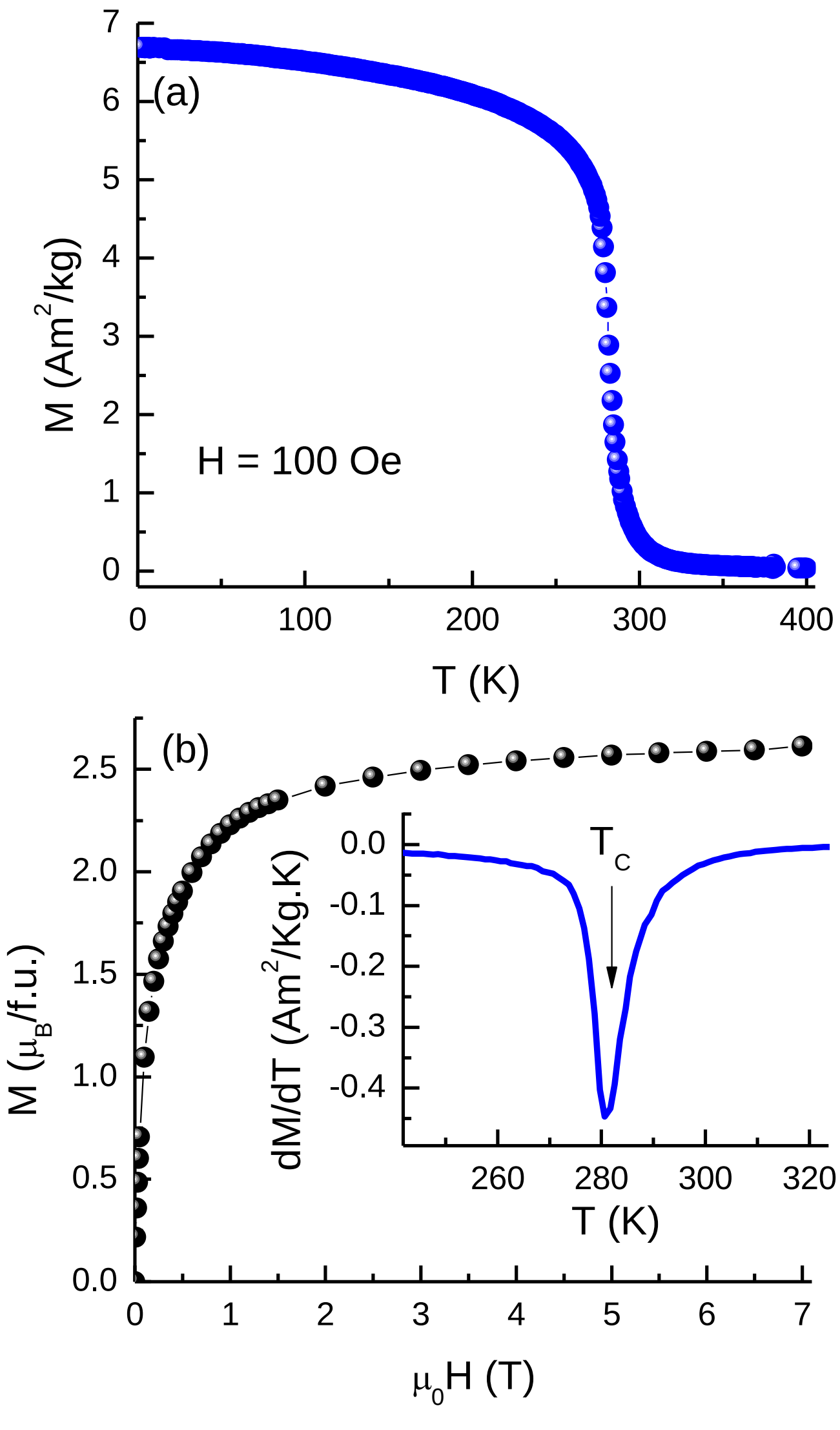

Fig. S2: (a) Temperature dependent zero field cooled magnetization behavior of the sample measured with 100 Oe. (b) M-*versus*-H behavior of sample measured at 2 K. The inset in (b) shows the first order derivative of magnetization with respect to temperature.

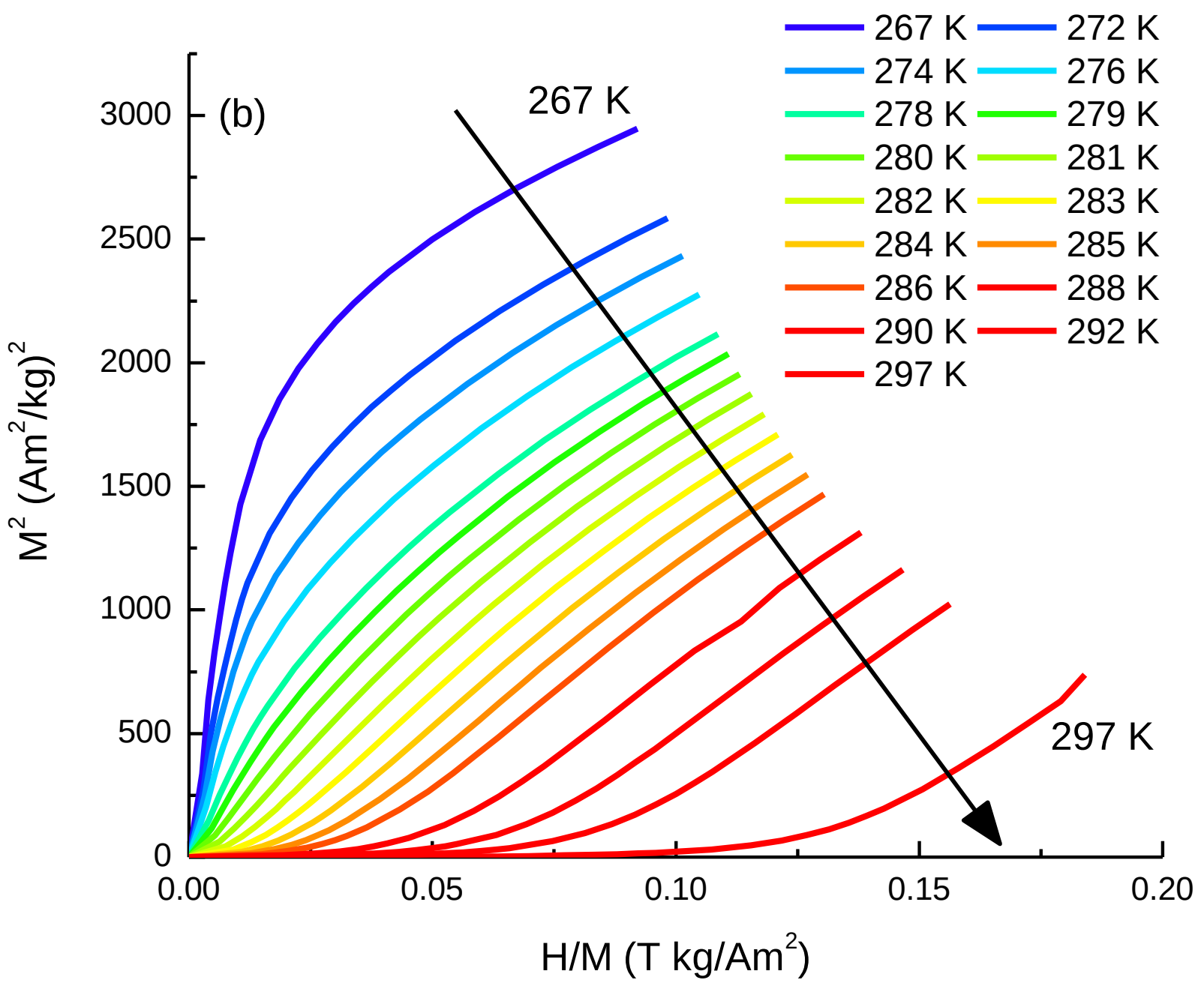

Fig. S3: Arrott plots measured at several temperatures across $T_C$ ranging from 267 to 297 K. The arrow indicates the direction of increasing temperature.

Interestingly, the comparative order of the effect of temperature and magnetic field on lattice parameters per degree and per tesla are almost the same. The values of the thermal coefficient per degree and per tesla along *c*-axis are 1.2E-5 $K^{-1}$ using the low temperature XRD data reported by Cedervall *et al.*[29] and 7.8E-5 $T^{-1}$ using the our experimental data at 290 K. The values of thermal coefficient per degree along *a* and *b*-axis are 9.4E-6 and 1.09E-5 $K^{-1}$ while the values per tesla are 2.32E-5 and 2.75E-5 $T^{-1}$ at 290 K.

Further, we have also estimated few components of the magnetostriction tensor based on the formalism of magnetoelastic energy term as mentioned in equation (1) of the manuscript. The magnetoelastic energy term has the form $\lambda\varepsilon M^2$. Magnetoelastic strain can be obtained by differentiating equation (1) with respect to strain $\varepsilon$ as shown in equation (2) of the manuscript. From equation (2), we can roughly estimate the N terms using the relation, $\varepsilon = NM^2$. For orthorhombic C*mmm* space group, the number of independent components in the magnetostriction tensor is 12 which cannot be obtained from the present powder data. However, it is feasible to roughly estimate a few components using the available data along the main crystallographic axes. In the magnetic structure of $AlFe_2B_2$, the moments are aligned along the $a$-axis[29] and we have magnetostriction mainly along $a$-, $b$- and $c$- axis. Hence using the equation $\varepsilon_a = N_{11}M^2, \varepsilon_b = N_{22}M^2$ and $\varepsilon_c = N_{33}M^2,$ we roughly estimate the three components $(N_{11}, N_{22}, N_{33})$ of the magnetoelastic tensor. The magnetic energy density at 25T is of the order of 248.7 MJ/m$^3$ and using the values of $\varepsilon_a, \varepsilon_b, \varepsilon_c$ from our experimental data, the value of $N_{11}, N_{22}, N_{33}$ are as follows: $N_{11}$ (along $a$-axis) = -2.1×10$^{-6}$ MPa$^{-1}$, $N_{22}$ (along $b$-axis) = -2.4×10$^{-6}$ MPa$^{-1}$, and $N_{33}$ (along $c$-axis) = 7.1×10$^{-6}$ MPa$^{-1}$.